\documentclass[conference]{IEEEtran}

\usepackage{rotating}

\IEEEoverridecommandlockouts
\usepackage{cite,xurl}
\usepackage{amsmath,amssymb,amsfonts}
\usepackage{algorithmic}
\usepackage{graphicx}
\usepackage{textcomp}
\usepackage{algorithm2e}
\usepackage{xcolor}
\def\BibTeX{{\rm B\kern-.05em{\sc i\kern-.025em b}\kern-.08em
    T\kern-.1667em\lower.7ex\hbox{E}\kern-.125emX}}

\usepackage{caption}
\usepackage{subcaption}
    
\usepackage{tikz}
\usetikzlibrary{quantikz}

\usepackage{subcaption}

\usepackage{amsmath}
\usepackage{amssymb}
\usepackage{amsthm}

\newtheoremstyle{ieeestyle}
  {\topsep}   
  {\topsep}   
  {\itshape}  
  {}          
  {\bfseries} 
  {.}         
  { }         
  {}          

\theoremstyle{ieeestyle}

\begin{document}

\title{Transversal Fanout for Fault Tolerant Distributed Quantum Computing: Analysis and Application}

\author{\IEEEauthorblockN{Seng W. Loke}
\IEEEauthorblockA{\textit{School of Information Technology, Deakin University, Burwood, VIC 3125, Australia.} \\
seng.loke@deakin.edu.au}
 }


\maketitle

\begin{abstract}
We study a resource-efficient approach for implementing logical fanout operations in fault-tolerant distributed quantum computing using transversal  operations on quantum error-correcting code blocks. Logical fanout, comprising multiple controlled-NOT operations from a common control qubit to target qubits located at remote nodes, is an important primitive for distributed quantum computation but can require substantial non-local communication when implemented directly between encoded blocks. We exploit the structure of encoded blocks and the availability of transversal logical operations to construct distributed fanout circuits that reduce the required non-local operations while preserving the logical action of the fanout operation. The construction is developed for encoded quantum information and illustrated using Bivariate-Bicycle (BB)-code blocks. We analyze the resulting physical gate, entanglement, circuit-depth, and ancilla requirements. The approach provides a systematic method for implementing large logical fanout operations across distributed error-corrected quantum processors. Also, we study a distributed implementation of the global gate $GCZ$ involving logical qubits (encoded using BB-code blocks), exploiting the concurrency in transversal distributed fanouts.  
\end{abstract}

\begin{IEEEkeywords}
distributed quantum computing, tansversal gate gadgets, fault-tolerant quantum computing, multipartite entanglement, distributed fanout gates
\end{IEEEkeywords}

\section{Introduction}

Distributed or modular quantum computing has been touted as a means of scaling up quantum computation, beyond  monolithic platform approaches~\cite{BARRAL2025100747,knörzer2025,CALEFFI2024110672,arquin}.\footnote{For example, see ~\cite{photonics}, \url{https://spectrum.ieee.org/quantum-computers}, \url{https://newsroom.ibm.com/2025-11-20-ibm-and-cisco-announce-plans-to-build-a-network-of-large-scale,-fault-tolerant-quantum-computers}. See also \url{https://quantnet.lbl.gov} and \url{https://www.ox.ac.uk/news/2025-02-06-first-distributed-quantum-algorithm-brings-quantum-supercomputers-closer}.} Much work has focused on the use of Bell pairs for realizing distributed or non-local (or inter-module) CNOT gates to connect disparate QPUs (Quantum Processing Units), as such gates together with single qubit gates can provide a universal gate set for quantum computations. There have indeed been many experimental realizations of such connectivity, see e.g., ~\cite{main25}.

Towards fault-tolerant distributed quantum computing, a range of encodings for quantum error correction  have been proposed and are being implemented, and even considered for distributed quantum computing including surface codes (e.g., ~\cite{singh2025modular,Chandra:2026ovs,ds45-fm9n}), Bivariate-Bicycle (BB) codes (e.g., transversal non-local CNOTs~\cite{Stack2026Transversal} and inter-module operations~\cite{yoder2025tour}), and floquet codes~\cite{sutcliffe2025distributed}.

Previous work advocated the use of {\em fan-out} operations, with one control qubit for multiple target qubits, amenable to efficient implementation in some types of quantum hardware such as trapped-ion quantum computers, enabling reduced depth quantum computations~\cite{fenner23,hoyer05}.  
Distributed fan-out using distributed GHZ states have been proposed in~\cite{Yimsiriwattana:2004xhy}, and investigated in~\cite{11662257,11500410}.

This paper discusses a 
transversal implementation of a distributed fanout with qubits in a Bivariate-Bicycle (BB) encoding, and also explore their use for realizing distributed versions of global quantum gates (or  
$GMS(\theta)$, short for Global Mølmer–Sørensen multi-qubit gates~\cite{PhysRevLett.82.1971}). 

In the rest of this paper, we first briefly review  distributed fan-out operations in \S2, and then look at distributed or non-local transversal fanout operations in \S3. In \S4, we provide a simulation analysis of distributed fanouts, and then explore applications of the transversal fanout for distributed versions of $GCZ$ global gates in \S5.   We  conclude with avenues for future work in \S6.

\section{Brief Review of Distributed Fan-Out Operations}

\begin{figure}
\centering
  \includegraphics[width=0.34\textwidth]{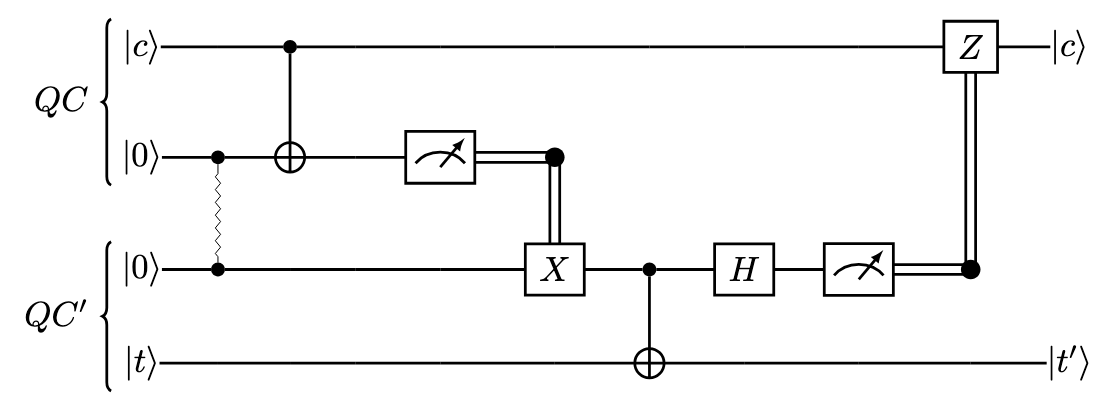}        $\equiv$ 
  \scalebox{0.8}{\begin{quantikz}  c~ & \ctrl{1}\gategroup[2,steps=1,style={dashed,inner sep=6pt}]{Dc-U}   & \qw  \\ 
t  & \gate{U}   & \qw    \end{quantikz} }
  \caption{Distributed controlled-$U$ (i.e., dCNOT is when U=$X$ ($\oplus$) as shown)   between nodes $QC$ and $QC'$ (the {\em computation} qubits are $c$, the control qubit, and $t$, the target qubit), and the wavy line illustrates a Bell pair  involving  the  {\em communication} qubits from the two nodes both initially $\ket{0}$. The right hand side shows the notation we will use for a distributed controlled-$U$ operation.}
\label{dcnot}
\end{figure}

We first consider the distributed CNOT gate (or  dCNOT, in short, a.k.a. non-local CNOT), which is needed as a resource for each dCNOT gate, as shown in Figure~\ref{dcnot}.

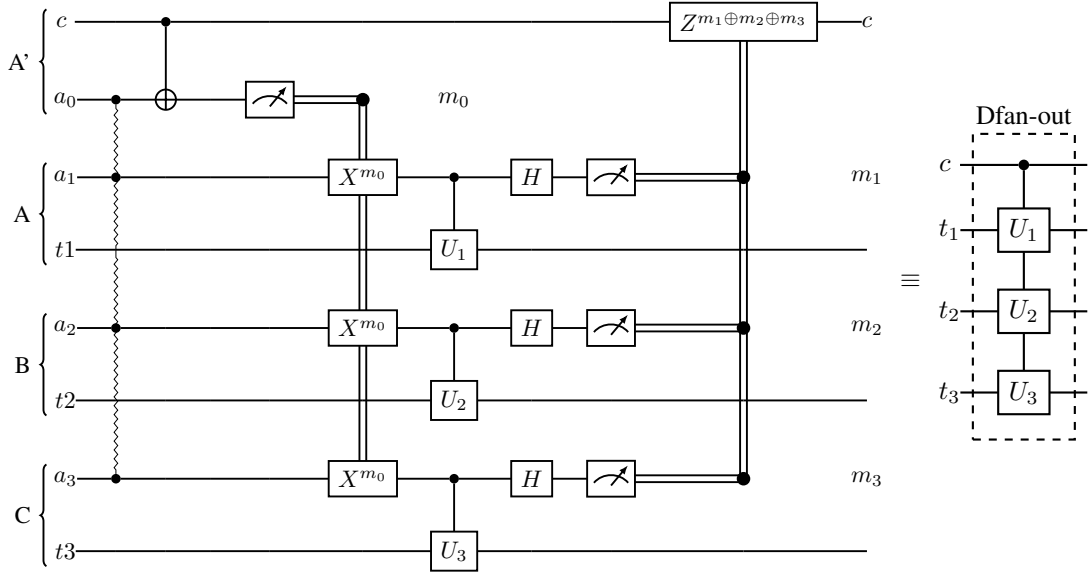
\begin{figure*}

\begin{center}
\scalebox{.9}{
      \begin{quantikz} 
   \lstick[2]{A'}    c~                           \qw   & \qw  & \ctrl{1}  & \qw   & \qw & \qw & \qw &   \qw &  \qw   &   \gate{Z^{m_1 \oplus m_2 \oplus m_3}} & \qw c  \\
    a_0    \qw & \ctrl{}{}  & \targ{}  & \qw & \meter{} &  \cwbend{5} &    m_0      &            &  &         \\
    \lstick[2]{A}    a_1    \qw & \ctrl{}{}  & \qw       & \qw & \qw      & \gate{X^{m_0}}   & \ctrl{1} & \gate{H}& \meter{} &  \cwbend{-2} &  m_1 \\
      t1   \qw & \qw        & \qw       & \qw & \qw      & \qw         & \gate{U_1}       & \qw        & \qw &  \qw &\qw   \\
          \lstick[2]{B}    a_2   \qw & \ctrl{}{}  & \qw       & \qw & \qw      & \gate{X^{m_0}}   & \ctrl{1} & \gate{H}  & \meter{} &  \cwbend{-2}   & m_2    \\
      t2   \qw & \qw        & \qw       & \qw & \qw      & \qw         & \gate{U_2}       & \qw        & \qw &  \qw &\qw  \\
                \lstick[2]{C}    a_3   \qw & \ctrl{}{}  & \qw       & \qw & \qw      & \gate{X^{m_0}}   & \ctrl{1} & \gate{H}  & \meter{} &  \cwbend{-2}   & m_3    \\
      t3   \qw & \qw        & \qw       & \qw & \qw      & \qw         & \gate{U_3}       & \qw        & \qw &  \qw &\qw  
         \arrow[from=2-2,to=4-2,squiggly,dash,line width=0.1mm]{}
         \arrow[from=7-2,to=4-2,squiggly,dash,line width=0.1mm]{}
    \end{quantikz}
    }
    $\equiv$
    \begin{quantikz}
 c~ & \ctrl{3}\gategroup[4,steps=1,style={dashed,inner sep=6pt}]{Dfan-out}   & \qw  \\
 t_1  & \gate{U_1}   & \qw \\ 
  t_2  & \gate{U_2}   & \qw \\ 
  t_3  & \gate{U_3}   & \qw 
\end{quantikz}
\end{center}
\caption{Distributed fanout operation with single control qubit (on $A'$) for multiple target qubits (one on $A$, one on $B$ and one on $C$) - all target qubits on different nodes from the control qubit. The right hand side shows the notation we will use for distributed fan-out throughout the paper. Note that $m_i \in \{0,1\}$.}
\label{dfanout}
\end{figure*}

By a distributed fanout  operation, we refer to an operation where the same control qubit is used for multiple target qubits, on different nodes. This can arise from the structure of a quantum circuit itself, or from a control-$U$ operation where the unitary $U$ spans multiple qubits and is decomposed into operations executed on qubits distributed on different nodes~\cite{Yimsiriwattana:2004xhy,loke2023distributed,PhysRevA.107.L060601}. For example, the quantum circuit such as:
    \begin{center}
          \begin{quantikz}
     c~     & \ctrl{1} & \ctrl{2}  & \ctrl{3}  & \qw       \\
     t_1     & \gate{U_1} &  \qw  & \qw  & \qw      \\
     t_2     & \qw        &  \gate{U_2}   & \qw   & \qw    \\
     t_3      & \qw     &  \qw  & \gate{U_3}  & \qw             \\
    \end{quantikz}
    ~$\equiv$~
    \begin{quantikz}
 c~ & \ctrl{3}\gategroup[4,steps=1,style={dashed,inner sep=6pt}]{fan-out}   & \qw  \\
 t_1  & \gate{U_1}   & \qw \\ 
  t_2  & \gate{U_2}   & \qw \\ 
  t_3  & \gate{U_3}   & \qw 
\end{quantikz}
    \end{center}
i.e., a {\em multitarget control-$U$},   where $U_1$ acts on qubit~$t1$, $U_2$ acts on qubit~$t2$, and $U_3$ acts on qubit~$t3$, each qubit on a different node, will result in the distributed operation as shown in Figure~\ref{dfanout}; note that the wavy line in the figure connecting the four (communication) qubits $a_0,a_1,a_2,a_3$  represents a distributed 4-qubit GHZ state of the form $\frac{1}{\sqrt{2}}(\ket{0 0 0 0}+\ket{1111})_{a_0 a_1 a_2 a_3}$, over four nodes $A'$, $A$, $B$ and $C$. 
In the rest of the paper, we particularly focus on the case where  $U_i = X$, unless otherwise noted.

\section{Transversal Non-Local Fanout}

We  outline why a non-local fanout operation can be implemented transversally.
A transversal operation can be defined as follows.
Let there be $m$ code blocks, each consisting of $n$ physical qubits. A quantum operation $U$ is transversal if it can be expressed as $U=\bigotimes_{i=1}^{n} U_i$,
where each $U_i$ acts only on the $i$-th physical qubits of the code blocks and does not couple two qubits within the same code block.

From~\cite{Stack2026Transversal}, we see that a logical CNOT  between two logical qubits is transversal, when the logical qubits are encoded using the Bivariate-Bicycle (BB) qLDPC code; multiple logical CNOT operations are performed simultaneously.
Because the logical fanout decomposes into CNOTs between the same code blocks, and because each of those logical CNOTs is transversal, their product can be reordered into a set of  independent physical fanout operations, one per physical coordinate; a logical fanout gadget is transversal if it can be composed using transversal logical CNOT gadgets. This comes from~\cite{Gottesman}, where ``a product of transversal gate gadgets is a transversal gate gadget'', and since, a logical fanout is a product of (transversal) CNOT operations, by definition.
Using the BB encoding $[[n,k,d]]$ of logical qubits, with operations on a block of $n$ qubits, we effectively perform $k$  independent logical fanout operations simultaneously.

 More specifically, using the $[[n,k,d]]$ $BB$ encoding, where the logical CNOT is transversal,  i.e., between two blocks $C$ with qubits $c_i$ and $T$ with qubits $t_i$, denoted by $\overline{CNOT}(C;T) = \Pi_{i\in \{1,...,n\}}CNOT(c_i;t_{i})$,
consider  one control block $C = \{c_1,...,c_n\}$ with $n$ qubits,  and $m$ target blocks $T_i$, $i \in \{1,...,m\}$, each with qubits $T_{i}=\{t_{i1},...,t_{in}\}$, where block $C$ encodes $k$ logical qubits $\{ \bar{c}^{(l)} ~|~ l \in \{1,...,k\}\}$ and each of the $m$ blocks, $T_i$, encodes $k$ logical qubits  $\{ \bar{t}^{(l)}_{i} ~|~ l \in \{1,...,k\} \}$. 
Suppose we wanted to perform the following operation:
\begin{align*}
 & \overline{FANOUT}(C;T_1,...,T_m) \\
:= ~~ & \bigotimes_{l \in \{1,...,k\}}  \overline{FANOUT}_L (\bar{c}^{(l)};\bar{t}^{(l)}_{1},...,\bar{t}^{(l)}_{m}) 
\end{align*} 
where $\overline{FANOUT}_L$   denotes the logical fanout acting on the corresponding $l$-th encoded logical qubit, $1 \leq l \leq k$. We do this by performing $n$ independent physical fanout operations, one for each coordinate of the code:
\begin{align*}
 &   \bigotimes_{i \in \{1,...,n\}} FANOUT(c_i;t_{1i},...,t_{mi})  
 \end{align*} 
 where $FANOUT$ denotes a physical fanout operation, each physical qubit $c_i$ fans out as control to the corresponding physical qubits in every target block.\footnote{We explain why this is so, as follows:
\begin{align*}
  &   \Pi_{i \in \{1,...,n\}} FANOUT(c_i;t_{1i},...,t_{mi})  \\
 = & \Pi_{i \in \{1,...,n\}}  \Pi_{j\in \{1,...,m\}}CNOT(c_i;t_{ji}) \\
 & \mbox{(from definition of $FANOUT$)} \\
 = & \Pi_{j \in \{1,...,m\}}  \Pi_{i\in \{1,...,n\}}CNOT(c_i;t_{ji}) \\
 & \mbox{(the required $CNOT$s commute)} \\
 = & \Pi_{j \in \{1,...,m\}}  \overline{CNOT}(C;T_j) \\
  & \mbox{(from transversal $CNOT$s between the control block $C$} \\
 & \mbox{and the target block $T_j$)} \\ 
 =  & \overline{FANOUT}(C;T_1,...,T_m)  
 \end{align*}

From another perspective, since a transversal CNOT between code blocks acts on  $k$ target logical qubits simultaneously, a transversal fanout among code blocks acts on  $k$ groups of target logical qubits simultaneously:
 \begin{align*}
   & \overline{FANOUT}(C;T_1,...,T_m)  \\
    = & \Pi_{j \in \{1,...,m\}}  \overline{CNOT}(C;T_j) \\
  = & \Pi_{j \in \{1,...,m\}}  \Pi_{l\in \{1,...,k\}} \overline{CNOT}_L(\bar{c}^{(l)};\bar{t}_j^{(l)}) \\
  = & \Pi_{l\in \{1,...,k\}} \Pi_{j \in \{1,...,m\}}  \overline{CNOT}_L(\bar{c}^{(l)};\bar{t}_j^{(l)}) \\
 & \mbox{(the required $\overline{CNOT}_L$s commute)} \\
=  & \Pi_{l \in \{1,...,k\}}  \overline{FANOUT}_L(\bar{c}^{(l)};\bar{t}^{(l)}_{1},...,\bar{t}^{(l)}_{m}) \\
 & \mbox{(from definition of $\overline{FANOUT}_L$)}  \\
 =  & \bigotimes_{l \in \{1,...,k\}}  \overline{FANOUT}_L(\bar{c}^{(l)};\bar{t}^{(l)}_{1},...,\bar{t}^{(l)}_{m})
\end{align*} 
}
Note that the above applies to non-local CNOTs and non-local FANOUTs, e.g., with one block per node (or $k$ logical qubits per node).

\subsection{Comparison with Pairwise Distributed CNOTs}

We can compare resources required for a GHZ-based fanout as shown in Figure~\ref{dfanout} with an equivalent (in terms of effect) sequence of CNOTs;
suppose one control (block on one node) simultaneously controls $m-1$ remote targets (i.e., $m$ nodes in total, one target block per target node, for simplicity), with comparison shown in Table~\ref{tab:pairwisevsghz}.
The GHZ based approach could  lead to a reduction in sources of errors, and reduced depth (which could also reduce decoherence due to reduced qubit wait times) and so, in lower error rates. However, it relies on being able to generate distributed GHZ states with low error rates.   We show later via simulation that this can work based on particular noise models.
With a distributed transversal fanout operation  among blocks, using BB block encoding $[[n,k,d]]$, with $n$ physical qubits, say for one block per node, over $m$ nodes,  we would have $n$ non-local physical $FANOUT$s, i.e., needing  $n$ $m$-qubit physical GHZ states. 
  For distributed GHZ states, a block of $n$ communication qubits are required on each node; this is similar for the distributed CNOTs approach with the communications qubits initialised and reused for each CNOT on the control node.
Table~\ref{tab:pairwisevsghz} summarises the comparison. The table treats an $m$-party GHZ as an available entanglement resource; the cost of generating that GHZ is architecture-dependent and is not included; similarly, with the Bell pairs.
 
\begin{table}[ht]
\centering
\caption{Physical resources (involving blocks of $n$ qubits)  for  distributed GHZ-based $\overline{FANOUT}$ and (equivalent) distributed  $\overline{CNOT}$s using $[[n,k,d]]BB$, $1$ control block and $m-1$ target blocks, over $m$ nodes (one block per node).}
\label{tab:pairwisevsghz}
\begin{tabular}{lcc}
\hline
\textbf{Resource} & \textbf{Distributed $\overline{CNOT}$s} & \textbf{GHZ $\overline{FANOUT}$} \\
\hline
Shared entanglement & $n(m-1)$ Bell pairs & $n$ m-party GHZ states \\
Local CNOTs & $2n(m-1)$ & $nm$ \\
Hadamards & $n(m-1)$ & $n(m-1)$ \\
Measurements & $2n(m-1)$ & $nm$ \\
Classical bits sent & $2n(m-1)$ &
$2n(m-1)$  \\
Communication qubits & $nm$ & $nm$ \\
\hline
\end{tabular}
\end{table}

\section{Simulation Study of Non-Local Fanouts}
\label{sec:simulation}
A fanout implements multiple CNOT interactions and may therefore expose the circuit to more error locations than a single CNOT; however, the GHZ-based implementation can reduce the number of non-local operations relative to a sequential realization. Given a physical error rate (PER) $p$ on qubits, and inter-node ebit noise  $p_{ebit}$, and noise in GHZ states $p_{ghz}$, one can compare the  LERs of a fanout operation using GHZ states and an equivalent sequence of CNOTs.   Also, a fanout operation  uses distributed GHZ states, each of which may be constructed using a linear number of distributed CNOTs, or one shot via other mechanisms (e.g., ~\cite{Ainley:2024bdu,6mqy-sd3d}) - both of which might have their own error sources and possibly yielding different LERs. 
Transversality establishes that the logical fanout can be implemented without coupling two physical qubits within the same code block. The circuit-level fault tolerance of the complete implementation, however, also depends on the syndrome-extraction circuit and decoder. Consequently, the simulations below evaluate the LER of the complete noisy circuit rather than assuming that the code distance is preserved automatically.

Here, we do comparisons, first  with a four node fanout operation, with one $[[36,4,6]] BB$ block per node, and then with eight nodes. Our simulation study is based on extending the Transversal Multiple Code Block Simulator (TMCBS) library code from~\cite{Stack2026Transversal} (created to study non-local transversal CNOTs) to simulate non-local fault-tolerant transversal fanout operations. TMCBS uses Stim~\cite{gidney2021stim}, and the syndrome extraction schedule used in the fanout simulation is given in Table~\ref{tab:syndrome-extraction}.    We utlised ChatGPT-5.6 Luna as a tool to generate Python implementations of the different versions of the fanout operations using the TMCBS library (and the Stim\footnote{\url{https://github.com/quantumlib/Stim}} library), in an interactive prompt and revise process, with some manual fine-tuning. We use the decoder Tesseract, like in~\cite{Stack2026Transversal}.

\begin{table*}[t]
\centering
\caption{Syndrome-extraction schedule (using $BB[[36,4,6]]$) used for the different distributed 
non-local gate implementations. Each initial extraction consists of one 
first-pass round followed by three repeated rounds. For 
\texttt{3cnot-seq}, an additional extraction round is performed 
between successive non-local CNOT operations; this isolates successive logical operations and provide the same fault-tolerance boundary used by the TMCBS non-local-CNOT protocol. No such intermediate extraction is required in the GHZ-fanout circuit because the fanout is treated as one logical operation.
DEPOLARIZE2 in Stim is a two-qubit depolarizing noise channel, with identity with probability $1-p_{ebit}$, and $p_{ebit}$, otherwise, i.e.
$Pr(II) =1-p_{ebit}, Pr(P) =\frac{p_{ebit}}{15}$, for $P\in\{IX,IY,IZ,XI,\ldots,ZZ\}$, and 
$\mathcal{E}_{\mathrm{GHZ}}(p_{\mathrm{ghz}}) =\prod_q \mathcal{D}_1(q,p_{\mathrm{ghz}}/2)$, where the implementation applies independent $\mathcal{D}_1$ (DEPOLARIZE1) channels to the GHZ qubits (where DEPOLARIZE1 is Stim’s single-qubit depolarizing channel) giving identity with probability $1-p_{\mathrm{ghz}}/2$ and $(p_{\mathrm{ghz}}/2)/3$ for $X$, $Y$ and $Z$ error, in this case. }
\label{tab:syndrome-extraction}
\begin{tabular}{lcccc}
\hline
\textbf{Circuit} &
\textbf{Before non-local operation} &
\textbf{Between operations} &
\textbf{After operation}  &
\textbf{Non-local noise model} \\\\
\hline
\texttt{1cnot}
& 4 rounds
& ---
& 3 rounds 
& DEPOLARIZE2 per dCNOT\\

\texttt{3cnot-seq}
& 4 rounds
& 1 round after each non-local CNOT
& 3 rounds 
& DEPOLARIZE2 per dCNOT \\

\texttt{3cnotghz-fanout}
& 4 rounds
& ---
& 3 rounds 
& DEPOLARIZE2 per dCNOT \\

\texttt{1shotghz-fanout}
& 4 rounds
& ---
& 3 rounds 
& $\mathcal{E}_{\mathrm{GHZ}}$ \\
\hline
\end{tabular}
\end{table*}

\begin{figure}
\centering
  \includegraphics[width=0.5\textwidth]{
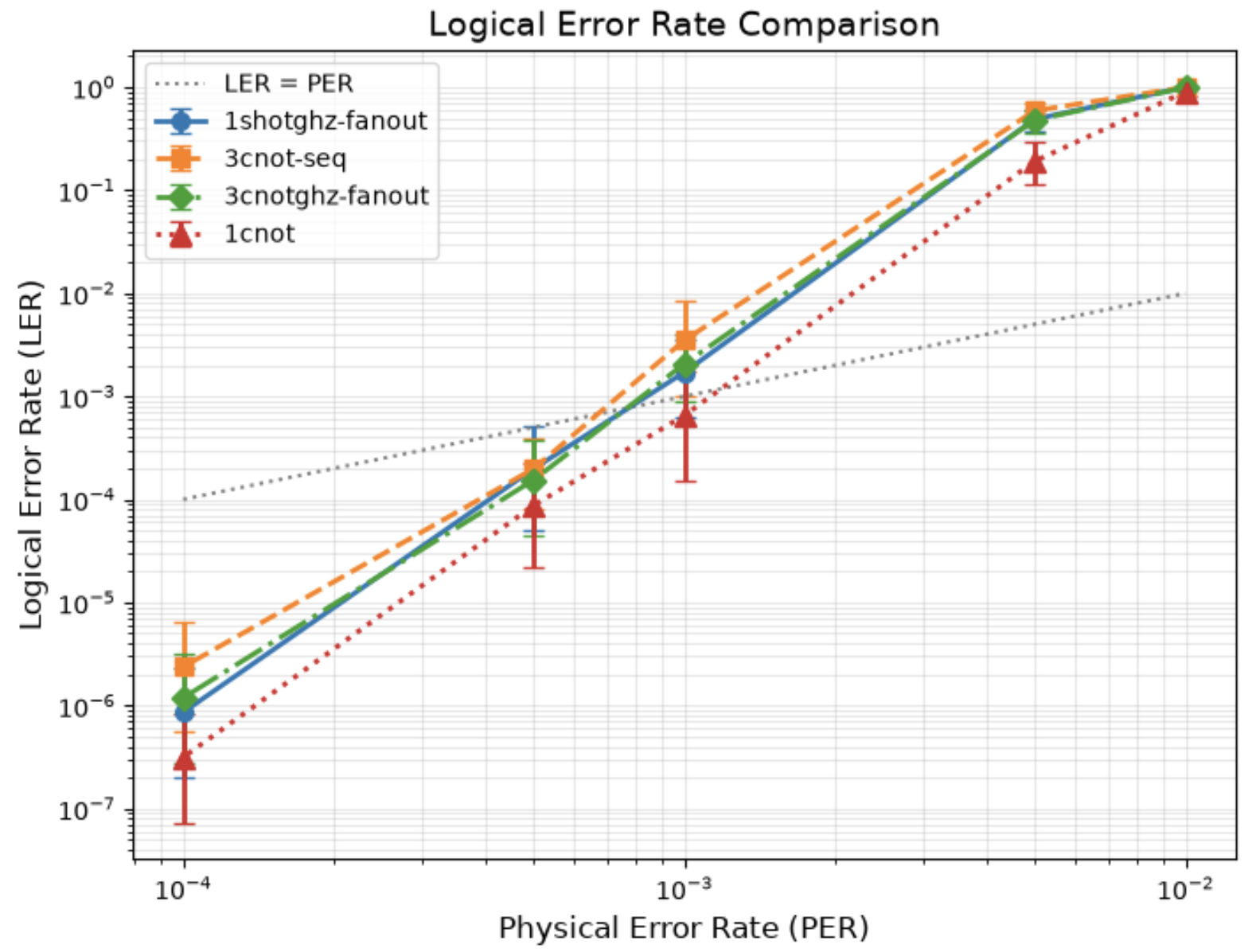} 
\centering
{\small 
\begin{tabular}{cccc}
\hline
$\mathbf{p}$ &
\textbf{1shotghz-fanout} &
\textbf{3cnotghz-fanout} &
\textbf{3cnot-seq} \\
\hline
$1\times10^{-2}$ & $1.00\times10^{0}$ & $9.92\times10^{-1}$ & $1.00\times10^{0}$ \\
$5\times10^{-3}$ & $4.88\times10^{-1}$ & $4.77\times10^{-1}$ & $5.90\times10^{-1}$ \\
$1\times10^{-3}$ & $1.71\times10^{-3}$ & $2.07\times10^{-3}$ & $3.50\times10^{-3}$ \\
$5\times10^{-4}$ & $1.98\times10^{-4}$ & $1.54\times10^{-4}$ & $1.98\times10^{-4}$ \\
$1\times10^{-4}$ & $8.60\times10^{-7}$ & $1.19\times10^{-6}$ & $2.37\times10^{-6}$ \\
\hline
\end{tabular}
}
  \caption{LER vs PER for a four node distributed fanout with $p=p_{ebit}=p_{ghz}$, comparing three different distributed fanout implementations - one using a one shot approach to create distributed GHZ states (1shotghz-fanout), one using GHZ states constructed using a sequence of neighbouring non-local CNOTs (3cnotghz-fanout), and one is a sequence of CNOTS (using only Bell pairs and not using GHZ states) equivalent to the fanout (3cnot-seq), as well as a single distributed CNOT (1cnot) for comparison. The table shows the plotted values.}
  \label{fig:4BB-LERvsPER}
\end{figure}

Figure~\ref{fig:4BB-LERvsPER}  shows the graph of LERs vs PERs for three simulated realizations of a distributed (or non-local) fanout, one using a one shot approach to create distributed GHZ states (1shotghz-fanout), one using GHZ states constructed using non-local CNOTs (3cnotghz-fanout), and one is a sequence of (non-local) CNOTS, equivalent to the fanout (3cnot-seq). We construct (blocks of) GHZ states for 1shotghz-fanout and 3cnotghz-fanout via a transversal protocol: among corresponding qubits in the communication qubit blocks on each node, perform the Hadamard $H$ on one ``root'' (control) qubit, and then perform non-local CNOT between the ``root'' and each of the other qubits. For the 1shotghz-fanout, we simulate this by constructing the GHZ state via the same protocol but counting the entire process as taking one simulation step. The respective noise models are shown in Table~\ref{tab:syndrome-extraction}.   
For each data point we also show error bars corresponding to 95\% confidence intervals.

In the first simulation,
we used $p=p_{ebit}=p_{ghz}$, representing physical error rate $p$ for a single qubit, the error rate for an $ebit$ and a parameter representing the physical error rate in a GHZ state (note that $p_{ghz}$ is not the probability of error in the GHZ state but a parameter to add errors to the state in the simulation as used in Table~\ref{tab:syndrome-extraction}), and repeated runs, and then calculating the LER, i.e.  the probability that at least one logical error occurs anywhere in the complete operation. Indeed, fanout using the 1-shot GHZ states approach generally provides the lowest LER in the low-PER regime, although the ordering of the two GHZ-based implementations can cross at intermediate error rates.  The table shows the cross-over when $LER < PER$,   when the PER $p \approx 5 \times 10^{-4}$ and below. 

\begin{figure}
\centering
  \includegraphics[width=0.5\textwidth]{
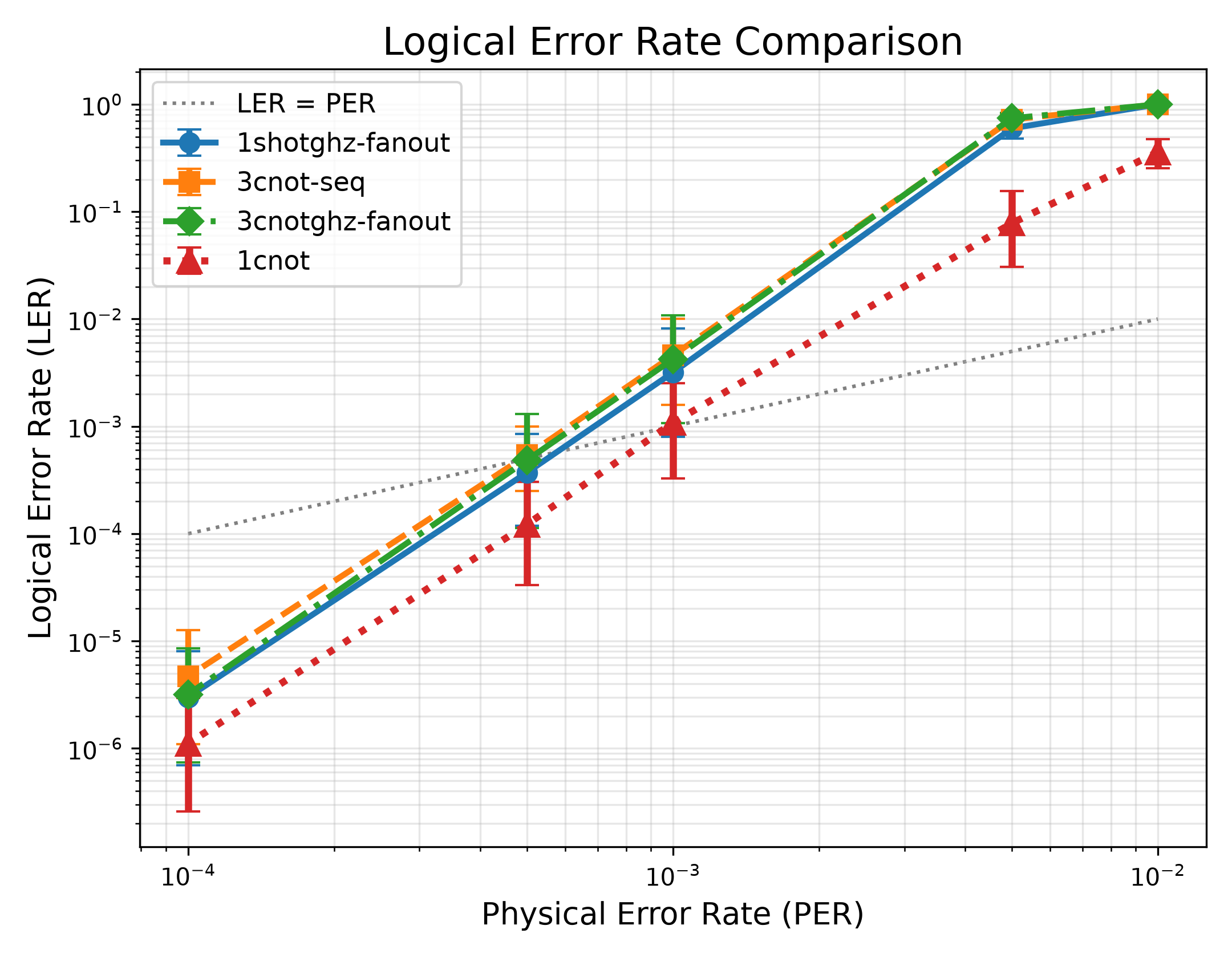} 
{\small 
\begin{tabular}{cccc}
\hline
$\mathbf{p}$ &
\textbf{1shotghz-fanout} &
\textbf{3cnotghz-fanout} &
\textbf{3cnot-seq} \\
\hline
$1\times10^{-2}$ & $1.00\times10^{0}$ & $1.00\times10^{0}$ & $1.00\times10^{0}$ \\
$5\times10^{-3}$ & $5.98\times10^{-1}$ & $7.42\times10^{-1}$ & $7.23\times10^{-1}$ \\
$1\times10^{-3}$ & $3.18\times10^{-3}$ & $4.23\times10^{-3}$ & $4.58\times10^{-3}$ \\
$5\times10^{-4}$ & $3.72\times10^{-4}$ & $4.85\times10^{-4}$ & $5.39\times10^{-4}$ \\
$1\times10^{-4}$ & $3.00\times10^{-6}$ & $3.19\times10^{-6}$ & $4.69\times10^{-6}$ \\
\hline
\end{tabular}
}
  \caption{LER vs PER for a four node distributed fanout with $10p=p_{ebit}=p_{ghz}$, comparing three different distributed fanout implementations - one using a one shot approach to create distributed GHZ states (1shotghz-fanout), one using GHZ states constructed using a sequence of neighbouring non-local CNOTs (3cnotghz-fanout), and one is a sequence of CNOTS (using only Bell pairs and not using GHZ states) equivalent to the fanout (3cnot-seq), as well as a single distributed CNOT (1cnot) for comparison. The table shows the plotted values.} 
  \label{fig:10p-4BB-LERvsPER}
\end{figure}

A corresponding set of results is shown in   Figure~\ref{fig:10p-4BB-LERvsPER}   when the physical error rate in the entanglement (distributed Bell pairs and  the GHZ states) is 10 times that of the PER on qubits, i.e.
$10p=p_{ebit}=p_{ghz}$; this is similar to the case of $p=p_{ebit}=p_{ghz}$, but with slightly larger differences between the fanout using 1-shot GHZ versus the other two approaches to fanout - higher entanglement costs makes the one shot GHZ approach relatively more efficient, as expected.

\begin{figure}
\centering
  \includegraphics[width=0.5\textwidth]{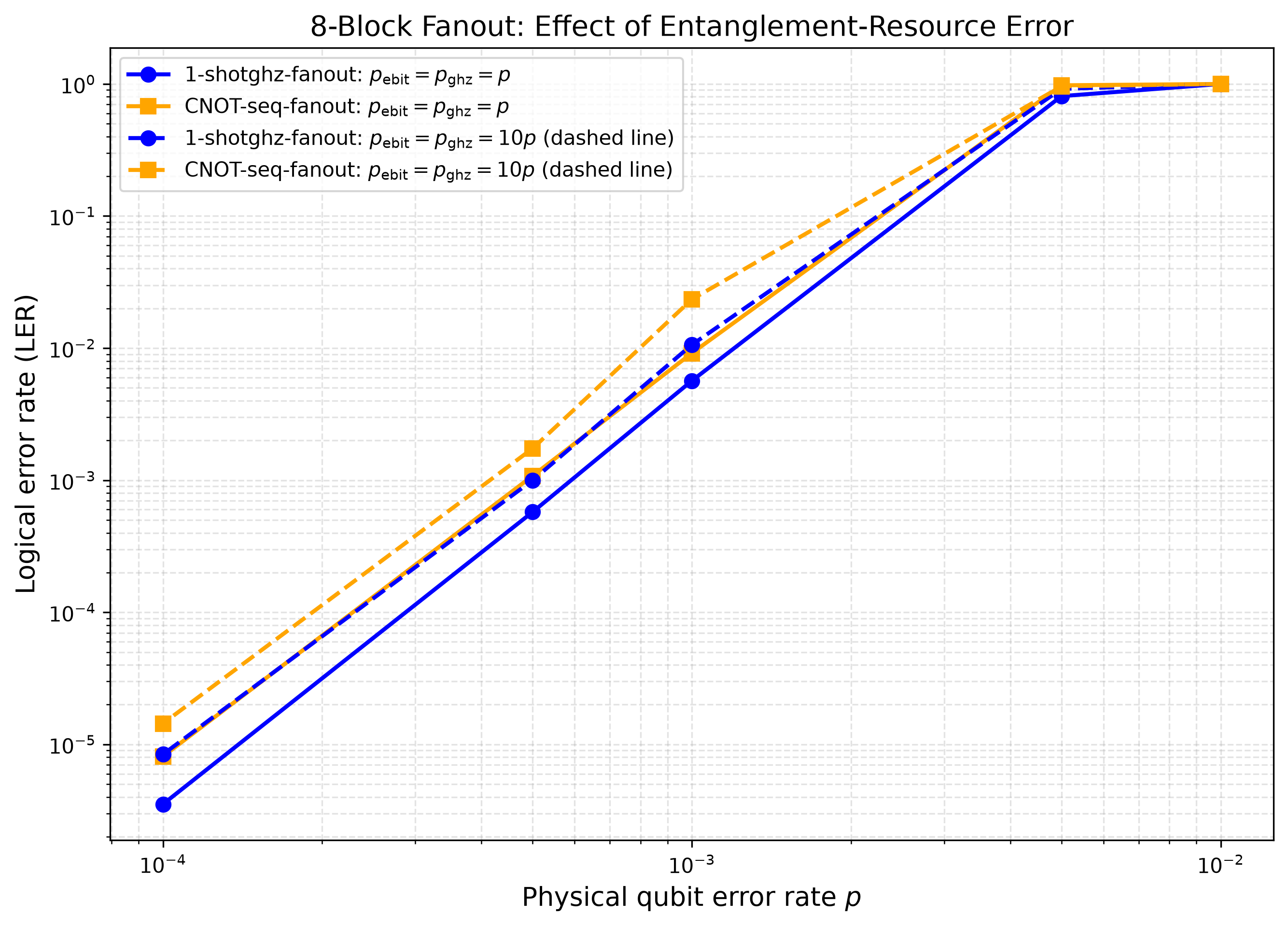} 
  \begin{tabular}{c|cc}
\hline
$p$ & 1shot-fanout & CNOT-seq fanout \\
\hline
$1\times10^{-2}$ & $1.00\times10^{0}$ & $1.00\times10^{0}$ \\
$5\times10^{-3}$ & $8.09\times10^{-1}$ & $9.77\times10^{-1}$ \\
$1\times10^{-3}$ & $5.66\times10^{-3}$ & $9.16\times10^{-3}$ \\
$5\times10^{-4}$ & $5.75\times10^{-4}$ & $1.08\times10^{-3}$ \\
$1\times10^{-4}$ & $3.53\times10^{-6}$ & $8.12\times10^{-6}$ \\
\hline
\end{tabular}
\begin{tabular}{c|cc} 
\hline
$p$ & 1shot-GHZ-fanout & CNOT-seq fanout \\
\hline
$1 \times 10^{-2}$ & $1.00 \times 10^{0}$ & $1.00 \times 10^{0}$ \\
$5 \times 10^{-3}$ & $9.18 \times 10^{-1}$ & $9.60 \times 10^{-1}$ \\
$1 \times 10^{-3}$ & $1.06 \times 10^{-2}$ & $2.34 \times 10^{-2}$ \\
$5 \times 10^{-4}$ & $1.00 \times 10^{-3}$ & $1.74 \times 10^{-3}$ \\
$1 \times 10^{-4}$ & $8.41 \times 10^{-6}$ & $1.44 \times 10^{-5}$ \\
\hline
\end{tabular}
  \caption{LER vs PER for an eight node (one data block per node) distributed fanout with $p=p_{ebit}=p_{ghz}$ (top table)  and $10p=p_{ebit}=p_{ghz}$ (bottom table), comparing two different distributed fanout implementations - one using a one shot approach to create distributed GHZ states (1shotghz) for the fanout,  and one is a sequence of CNOTS (using only Bell pairs) equivalent to the fanout (CNOT-seq).} 
   \label{fig:10pandp-8BB-LERvsPER}
\end{figure}

\begin{figure}
\centering
  \includegraphics[width=0.5\textwidth]{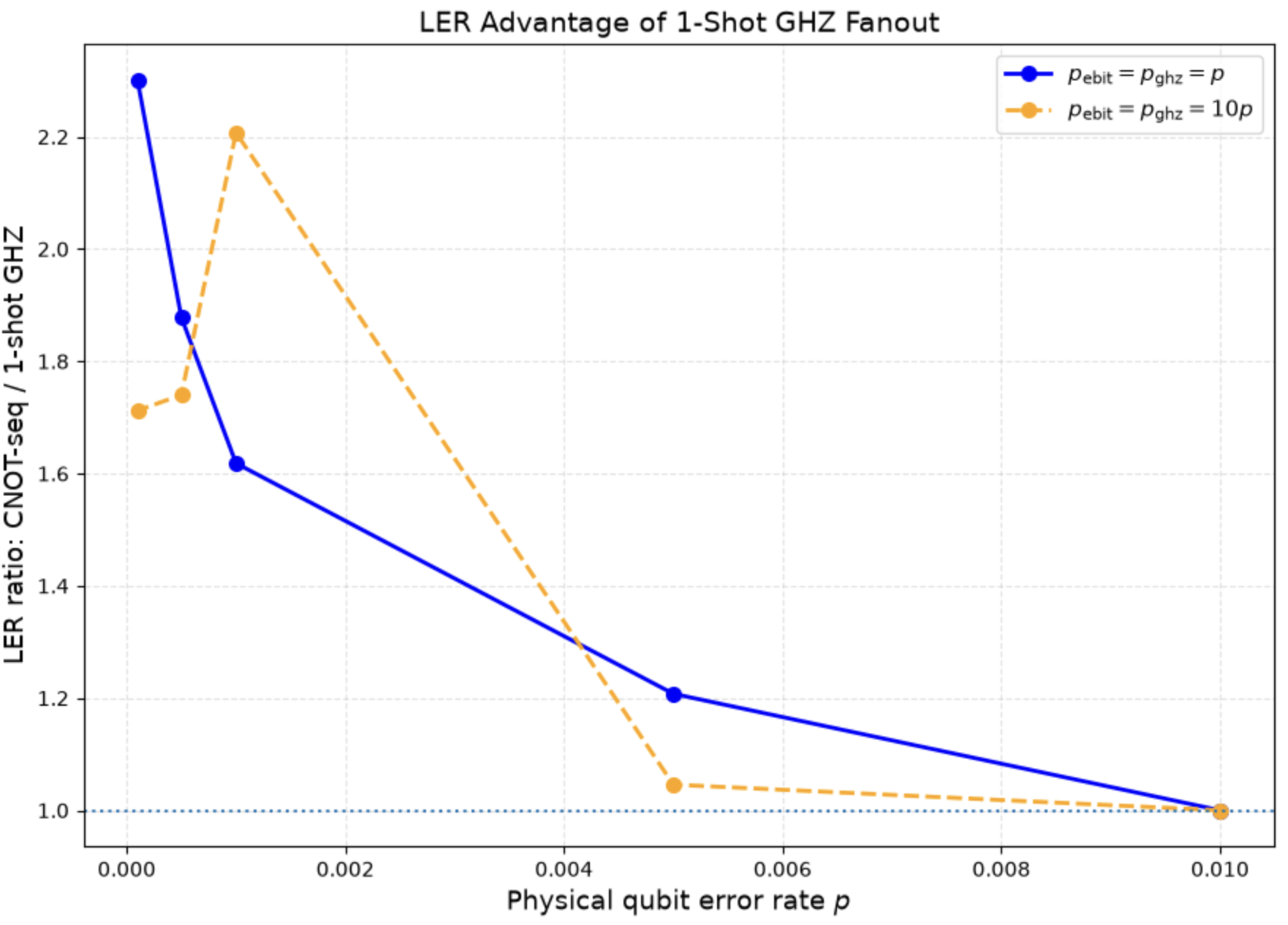} 
  \begin{tabular}{c|cc}
\hline
$p$ &
$\displaystyle p_{\mathrm{ebit}}=p_{\mathrm{ghz}}=p$ &
$\displaystyle p_{\mathrm{ebit}}=p_{\mathrm{ghz}}=10p$ \\
\hline
$1\times10^{-2}$ & $1.00$ & $1.00$ \\
$5\times10^{-3}$ & $1.21$ & $1.05$ \\
$1\times10^{-3}$ & $1.62$ & $2.21$ \\
$5\times10^{-4}$ & $1.88$ & $1.74$ \\
$1\times10^{-4}$ & $2.30$ & $1.71$ \\
\hline
\end{tabular}
  \caption{LER improvement for $10p=p_{ebit}=p_{ghz}$ and $p=p_{ebit}=p_{ghz}$. LER improvement factor of the 1-shot GHZ fanout relative to the
sequential CNOT fanout for 8 blocks. The improvement factor is defined as $\mathrm{LER}_{\mathrm{CNOT\text{-}seq}}/
 \mathrm{LER}_{\mathrm{1shotghz}}$.}
  \label{fig:ler-improvement-8block}
\end{figure}

We also performed similar comparisons with distributed fanout operations over eight nodes, shown in Figure~\ref{fig:10pandp-8BB-LERvsPER} with the values shown in the tables.
As expected, LER is higher for 8 nodes, compared to 4 nodes, and the 1-shot GHZ fanout has lower LER.  The improvement factor using the 1-shot GHZ fanout compared to using just Bell pairs (a sequence of CNOTs) for the 8 block (or 8 nodes) distributed fanout is shown in Figure~\ref{fig:ler-improvement-8block} - up to around a factor of 2.3 can be achieved in this case with low $p$.
Overall, even with the $[[36,4,6]]$, for all cases, a LER in the order of $\approx 10^{-6}$, below PER, is achieved for $p=10^{-4}$.

We studied the circuit-level distance  of   the non-local transversal fanout circuits using both a sequence of non-local CNOTs and using a 1-shot GHZ state fanout approach    via the STIM undetectable-error search heuristic (i.e., \texttt{search\_for\_undetectable\_logical\_errors}), like in ~\cite{Stack2026Transversal}. Code distance refers to the minimum weight of a Pauli operator implementing a logical operation; given a circuit $circ$, $d_{circ}$ refers to the minimum number of errors in the logical circuit (across time and space),   causing an unwanted logical operation on an observable (i.e., a logical qubit).  Table~\ref{tab:circdistance} shows the results; the search found no undetectable logical error of fault weight below 6, i.e., $d_{circ}=6$  for both 4 and 8 node fanouts.

\begin{table}[t]
\centering
\caption{Circuit-level distances for BB-code distributed fanout circuits using a sequence of (non-local) CNOTs and using 1-shot distributed GHZ state fanout, using $[[36,4,6]] BB$ encoding for the logical qubits, one block per node - the results are the same for both 4 and 8 nodes; included is $d_{circ}$ for the non-local CNOT, for comparison,  from~\cite{Stack2026Transversal}.}
\label{tab:circdistance}
\begin{tabular}{c c c c}
\hline
BB Code & $d$ & Circuit & $d_{\mathrm{circ}}$   \\
\hline
$[[36,4,6]]$
    & 6 & Non-local CNOT
    & 6   \\
$[[36,4,6]]$
    & 6 & CNOT sequence fanout
    & 6  \\
$[[36,4,6]]$
    & 6 & 1-shot GHZ fanout
    & 6 \\
\hline
\end{tabular}
\end{table}

\section{Distributed GCZ - an Application of Transversal Non-Local Fanouts}
We have seen that each transversal fanout operation performs $k$ logical fanout operations concurrently since the $n$ qubits of each block encodes $k$ logical qubits. This can be advantageous for operations involving multiple fanouts, such as the so-called   global gates  - this section illustrates this. 

Global gates   are  efficient for some quantum architectures such as trapped ion qubits where entire arrays of qubits can  be targeted and pairwise qubit-qubit interactions   can be naturally realized, which can be efficient for certain applications~\cite{Wetering2021,maslovandnam2018}.
A global MS gate or operation (GMS) over a set of qubits $S$ is defined as follows: $GMS_S(\theta)=$
\[
 exp(-i\frac{\theta}{2} \sum_{i,j\in S, ~ i <j} X_i X_j)  = \prod_{i,j\in S, ~ i <j} exp(-i\frac{\theta}{2}X_i X_j)
 \]
where   $exp(-i\frac{\theta}{2}X_i X_j)$ is called the ``local'' MS gate acting on qubits $i$ and $j$, and can be viewed operationally as: 
$exp(-i\frac{\theta}{2}X_i X_j)$ = 
$(H_i \otimes H_j)  CNOT_{i\rightarrow j}  (I_i \otimes (R_Z(\theta))_j)  CNOT_{i\rightarrow j} (H_i \otimes H_j)$,
with  $R_Z(\theta) = \begin{bmatrix} e^{-i\frac{\theta}{2}} & 0 \\ 0 & e^{i\frac{\theta}{2}}  \end{bmatrix}$.
(Note that the local MS gate (or LMS gate, for short) is synmmetrical, i.e. $exp(-i\frac{\theta}{2}X_i X_j)$=$exp(-i\frac{\theta}{2}X_j X_i)$.) GCZ gates are equivalent (up to Clifford gates) to $GMS(\pi/2)$ gates~\cite{Wetering2021}.

An illustration of a $GCZ_{16}$ operation involving 16 (logical) qubits over four nodes, four (logical) qubits per node, is shown in Figure~\ref{fig:gcz120}, where there are 120 CZ operations, of which $16(16-1)/2 - 4*4(4-1)/2 = 96$ CZs are inter-node CZs.

\begin{figure*}[t]
    \centering
    \resizebox{\linewidth}{!}{%
    \begin{quantikz}
\lstick{\text{Node 1: }$q_1$} & \ctrl{1} & \ctrl{2} & \ctrl{3} & \ctrl{4} & \ctrl{5} & \ctrl{6} & \ctrl{7} & \ctrl{8} & \ctrl{9} & \ctrl{10} & \ctrl{11} & \ctrl{12} & \ctrl{13} & \ctrl{14} & \ctrl{15} & \qw & \qw & \qw & \qw & \qw & \qw & \qw & \qw & \qw & \qw & \qw & \qw & \qw & \qw & \qw & \qw & \qw & \qw & \qw & \qw & \qw & \qw & \qw & \qw & \qw & \qw & \qw & \qw & \qw & \qw & \qw & \qw & \qw & \qw & \qw & \qw & \qw & \qw & \qw & \qw & \qw & \qw & \qw & \qw & \qw & \qw & \qw & \qw & \qw & \qw & \qw & \qw & \qw & \qw & \qw & \qw & \qw & \qw & \qw & \qw & \qw & \qw & \qw & \qw & \qw & \qw & \qw & \qw & \qw & \qw & \qw & \qw & \qw & \qw & \qw & \qw & \qw & \qw & \qw & \qw & \qw & \qw & \qw & \qw & \qw & \qw & \qw & \qw & \qw & \qw & \qw & \qw & \qw & \qw & \qw & \qw & \qw & \qw & \qw & \qw & \qw & \qw & \qw & \qw & \qw & \qw \\
\lstick{$q_{2}$} & \control{} & \qw & \qw & \qw & \qw & \qw & \qw & \qw & \qw & \qw & \qw & \qw & \qw & \qw & \qw & \ctrl{1} & \ctrl{2} & \ctrl{3} & \ctrl{4} & \ctrl{5} & \ctrl{6} & \ctrl{7} & \ctrl{8} & \ctrl{9} & \ctrl{10} & \ctrl{11} & \ctrl{12} & \ctrl{13} & \ctrl{14} & \qw & \qw & \qw & \qw & \qw & \qw & \qw & \qw & \qw & \qw & \qw & \qw & \qw & \qw & \qw & \qw & \qw & \qw & \qw & \qw & \qw & \qw & \qw & \qw & \qw & \qw & \qw & \qw & \qw & \qw & \qw & \qw & \qw & \qw & \qw & \qw & \qw & \qw & \qw & \qw & \qw & \qw & \qw & \qw & \qw & \qw & \qw & \qw & \qw & \qw & \qw & \qw & \qw & \qw & \qw & \qw & \qw & \qw & \qw & \qw & \qw & \qw & \qw & \qw & \qw & \qw & \qw & \qw & \qw & \qw & \qw & \qw & \qw & \qw & \qw & \qw & \qw & \qw & \qw & \qw & \qw & \qw & \qw & \qw & \qw & \qw & \qw & \qw & \qw & \qw & \qw & \qw \\
\lstick{$q_{3}$} & \qw & \control{} & \qw & \qw & \qw & \qw & \qw & \qw & \qw & \qw & \qw & \qw & \qw & \qw & \qw & \control{} & \qw & \qw & \qw & \qw & \qw & \qw & \qw & \qw & \qw & \qw & \qw & \qw & \qw & \ctrl{1} & \ctrl{2} & \ctrl{3} & \ctrl{4} & \ctrl{5} & \ctrl{6} & \ctrl{7} & \ctrl{8} & \ctrl{9} & \ctrl{10} & \ctrl{11} & \ctrl{12} & \ctrl{13} & \qw & \qw & \qw & \qw & \qw & \qw & \qw & \qw & \qw & \qw & \qw & \qw & \qw & \qw & \qw & \qw & \qw & \qw & \qw & \qw & \qw & \qw & \qw & \qw & \qw & \qw & \qw & \qw & \qw & \qw & \qw & \qw & \qw & \qw & \qw & \qw & \qw & \qw & \qw & \qw & \qw & \qw & \qw & \qw & \qw & \qw & \qw & \qw & \qw & \qw & \qw & \qw & \qw & \qw & \qw & \qw & \qw & \qw & \qw & \qw & \qw & \qw & \qw & \qw & \qw & \qw & \qw & \qw & \qw & \qw & \qw & \qw & \qw & \qw & \qw & \qw & \qw & \qw & \qw \\
\lstick{$q_{4}$} & \qw & \qw & \control{} & \qw & \qw & \qw & \qw & \qw & \qw & \qw & \qw & \qw & \qw & \qw & \qw & \qw & \control{} & \qw & \qw & \qw & \qw & \qw & \qw & \qw & \qw & \qw & \qw & \qw & \qw & \control{} & \qw & \qw & \qw & \qw & \qw & \qw & \qw & \qw & \qw & \qw & \qw & \qw & \ctrl{1} & \ctrl{2} & \ctrl{3} & \ctrl{4} & \ctrl{5} & \ctrl{6} & \ctrl{7} & \ctrl{8} & \ctrl{9} & \ctrl{10} & \ctrl{11} & \ctrl{12} & \qw & \qw & \qw & \qw & \qw & \qw & \qw & \qw & \qw & \qw & \qw & \qw & \qw & \qw & \qw & \qw & \qw & \qw & \qw & \qw & \qw & \qw & \qw & \qw & \qw & \qw & \qw & \qw & \qw & \qw & \qw & \qw & \qw & \qw & \qw & \qw & \qw & \qw & \qw & \qw & \qw & \qw & \qw & \qw & \qw & \qw & \qw & \qw & \qw & \qw & \qw & \qw & \qw & \qw & \qw & \qw & \qw & \qw & \qw & \qw & \qw & \qw & \qw & \qw & \qw & \qw & \qw \\
\lstick{\text{Node 2: }$q_5$} & \qw & \qw & \qw & \control{} & \qw & \qw & \qw & \qw & \qw & \qw & \qw & \qw & \qw & \qw & \qw & \qw & \qw & \control{} & \qw & \qw & \qw & \qw & \qw & \qw & \qw & \qw & \qw & \qw & \qw & \qw & \control{} & \qw & \qw & \qw & \qw & \qw & \qw & \qw & \qw & \qw & \qw & \qw & \control{} & \qw & \qw & \qw & \qw & \qw & \qw & \qw & \qw & \qw & \qw & \qw & \ctrl{1} & \ctrl{2} & \ctrl{3} & \ctrl{4} & \ctrl{5} & \ctrl{6} & \ctrl{7} & \ctrl{8} & \ctrl{9} & \ctrl{10} & \ctrl{11} & \qw & \qw & \qw & \qw & \qw & \qw & \qw & \qw & \qw & \qw & \qw & \qw & \qw & \qw & \qw & \qw & \qw & \qw & \qw & \qw & \qw & \qw & \qw & \qw & \qw & \qw & \qw & \qw & \qw & \qw & \qw & \qw & \qw & \qw & \qw & \qw & \qw & \qw & \qw & \qw & \qw & \qw & \qw & \qw & \qw & \qw & \qw & \qw & \qw & \qw & \qw & \qw & \qw & \qw & \qw & \qw \\
\lstick{$q_{6}$} & \qw & \qw & \qw & \qw & \control{} & \qw & \qw & \qw & \qw & \qw & \qw & \qw & \qw & \qw & \qw & \qw & \qw & \qw & \control{} & \qw & \qw & \qw & \qw & \qw & \qw & \qw & \qw & \qw & \qw & \qw & \qw & \control{} & \qw & \qw & \qw & \qw & \qw & \qw & \qw & \qw & \qw & \qw & \qw & \control{} & \qw & \qw & \qw & \qw & \qw & \qw & \qw & \qw & \qw & \qw & \control{} & \qw & \qw & \qw & \qw & \qw & \qw & \qw & \qw & \qw & \qw & \ctrl{1} & \ctrl{2} & \ctrl{3} & \ctrl{4} & \ctrl{5} & \ctrl{6} & \ctrl{7} & \ctrl{8} & \ctrl{9} & \ctrl{10} & \qw & \qw & \qw & \qw & \qw & \qw & \qw & \qw & \qw & \qw & \qw & \qw & \qw & \qw & \qw & \qw & \qw & \qw & \qw & \qw & \qw & \qw & \qw & \qw & \qw & \qw & \qw & \qw & \qw & \qw & \qw & \qw & \qw & \qw & \qw & \qw & \qw & \qw & \qw & \qw & \qw & \qw & \qw & \qw & \qw & \qw \\
\lstick{$q_{7}$} & \qw & \qw & \qw & \qw & \qw & \control{} & \qw & \qw & \qw & \qw & \qw & \qw & \qw & \qw & \qw & \qw & \qw & \qw & \qw & \control{} & \qw & \qw & \qw & \qw & \qw & \qw & \qw & \qw & \qw & \qw & \qw & \qw & \control{} & \qw & \qw & \qw & \qw & \qw & \qw & \qw & \qw & \qw & \qw & \qw & \control{} & \qw & \qw & \qw & \qw & \qw & \qw & \qw & \qw & \qw & \qw & \control{} & \qw & \qw & \qw & \qw & \qw & \qw & \qw & \qw & \qw & \control{} & \qw & \qw & \qw & \qw & \qw & \qw & \qw & \qw & \qw & \ctrl{1} & \ctrl{2} & \ctrl{3} & \ctrl{4} & \ctrl{5} & \ctrl{6} & \ctrl{7} & \ctrl{8} & \ctrl{9} & \qw & \qw & \qw & \qw & \qw & \qw & \qw & \qw & \qw & \qw & \qw & \qw & \qw & \qw & \qw & \qw & \qw & \qw & \qw & \qw & \qw & \qw & \qw & \qw & \qw & \qw & \qw & \qw & \qw & \qw & \qw & \qw & \qw & \qw & \qw & \qw & \qw \\
\lstick{$q_{8}$} & \qw & \qw & \qw & \qw & \qw & \qw & \control{} & \qw & \qw & \qw & \qw & \qw & \qw & \qw & \qw & \qw & \qw & \qw & \qw & \qw & \control{} & \qw & \qw & \qw & \qw & \qw & \qw & \qw & \qw & \qw & \qw & \qw & \qw & \control{} & \qw & \qw & \qw & \qw & \qw & \qw & \qw & \qw & \qw & \qw & \qw & \control{} & \qw & \qw & \qw & \qw & \qw & \qw & \qw & \qw & \qw & \qw & \control{} & \qw & \qw & \qw & \qw & \qw & \qw & \qw & \qw & \qw & \control{} & \qw & \qw & \qw & \qw & \qw & \qw & \qw & \qw & \control{} & \qw & \qw & \qw & \qw & \qw & \qw & \qw & \qw & \ctrl{1} & \ctrl{2} & \ctrl{3} & \ctrl{4} & \ctrl{5} & \ctrl{6} & \ctrl{7} & \ctrl{8} & \qw & \qw & \qw & \qw & \qw & \qw & \qw & \qw & \qw & \qw & \qw & \qw & \qw & \qw & \qw & \qw & \qw & \qw & \qw & \qw & \qw & \qw & \qw & \qw & \qw & \qw & \qw & \qw & \qw \\
\lstick{\text{Node 3: }$q_9$} & \qw & \qw & \qw & \qw & \qw & \qw & \qw & \control{} & \qw & \qw & \qw & \qw & \qw & \qw & \qw & \qw & \qw & \qw & \qw & \qw & \qw & \control{} & \qw & \qw & \qw & \qw & \qw & \qw & \qw & \qw & \qw & \qw & \qw & \qw & \control{} & \qw & \qw & \qw & \qw & \qw & \qw & \qw & \qw & \qw & \qw & \qw & \control{} & \qw & \qw & \qw & \qw & \qw & \qw & \qw & \qw & \qw & \qw & \control{} & \qw & \qw & \qw & \qw & \qw & \qw & \qw & \qw & \qw & \control{} & \qw & \qw & \qw & \qw & \qw & \qw & \qw & \qw & \control{} & \qw & \qw & \qw & \qw & \qw & \qw & \qw & \control{} & \qw & \qw & \qw & \qw & \qw & \qw & \qw & \ctrl{1} & \ctrl{2} & \ctrl{3} & \ctrl{4} & \ctrl{5} & \ctrl{6} & \ctrl{7} & \qw & \qw & \qw & \qw & \qw & \qw & \qw & \qw & \qw & \qw & \qw & \qw & \qw & \qw & \qw & \qw & \qw & \qw & \qw & \qw & \qw & \qw \\
\lstick{$q_{10}$} & \qw & \qw & \qw & \qw & \qw & \qw & \qw & \qw & \control{} & \qw & \qw & \qw & \qw & \qw & \qw & \qw & \qw & \qw & \qw & \qw & \qw & \qw & \control{} & \qw & \qw & \qw & \qw & \qw & \qw & \qw & \qw & \qw & \qw & \qw & \qw & \control{} & \qw & \qw & \qw & \qw & \qw & \qw & \qw & \qw & \qw & \qw & \qw & \control{} & \qw & \qw & \qw & \qw & \qw & \qw & \qw & \qw & \qw & \qw & \control{} & \qw & \qw & \qw & \qw & \qw & \qw & \qw & \qw & \qw & \control{} & \qw & \qw & \qw & \qw & \qw & \qw & \qw & \qw & \control{} & \qw & \qw & \qw & \qw & \qw & \qw & \qw & \control{} & \qw & \qw & \qw & \qw & \qw & \qw & \control{} & \qw & \qw & \qw & \qw & \qw & \qw & \ctrl{1} & \ctrl{2} & \ctrl{3} & \ctrl{4} & \ctrl{5} & \ctrl{6} & \qw & \qw & \qw & \qw & \qw & \qw & \qw & \qw & \qw & \qw & \qw & \qw & \qw & \qw & \qw & \qw \\
\lstick{$q_{11}$} & \qw & \qw & \qw & \qw & \qw & \qw & \qw & \qw & \qw & \control{} & \qw & \qw & \qw & \qw & \qw & \qw & \qw & \qw & \qw & \qw & \qw & \qw & \qw & \control{} & \qw & \qw & \qw & \qw & \qw & \qw & \qw & \qw & \qw & \qw & \qw & \qw & \control{} & \qw & \qw & \qw & \qw & \qw & \qw & \qw & \qw & \qw & \qw & \qw & \control{} & \qw & \qw & \qw & \qw & \qw & \qw & \qw & \qw & \qw & \qw & \control{} & \qw & \qw & \qw & \qw & \qw & \qw & \qw & \qw & \qw & \control{} & \qw & \qw & \qw & \qw & \qw & \qw & \qw & \qw & \control{} & \qw & \qw & \qw & \qw & \qw & \qw & \qw & \control{} & \qw & \qw & \qw & \qw & \qw & \qw & \control{} & \qw & \qw & \qw & \qw & \qw & \control{} & \qw & \qw & \qw & \qw & \qw & \ctrl{1} & \ctrl{2} & \ctrl{3} & \ctrl{4} & \ctrl{5} & \qw & \qw & \qw & \qw & \qw & \qw & \qw & \qw & \qw & \qw & \qw \\
\lstick{$q_{12}$} & \qw & \qw & \qw & \qw & \qw & \qw & \qw & \qw & \qw & \qw & \control{} & \qw & \qw & \qw & \qw & \qw & \qw & \qw & \qw & \qw & \qw & \qw & \qw & \qw & \control{} & \qw & \qw & \qw & \qw & \qw & \qw & \qw & \qw & \qw & \qw & \qw & \qw & \control{} & \qw & \qw & \qw & \qw & \qw & \qw & \qw & \qw & \qw & \qw & \qw & \control{} & \qw & \qw & \qw & \qw & \qw & \qw & \qw & \qw & \qw & \qw & \control{} & \qw & \qw & \qw & \qw & \qw & \qw & \qw & \qw & \qw & \control{} & \qw & \qw & \qw & \qw & \qw & \qw & \qw & \qw & \control{} & \qw & \qw & \qw & \qw & \qw & \qw & \qw & \control{} & \qw & \qw & \qw & \qw & \qw & \qw & \control{} & \qw & \qw & \qw & \qw & \qw & \control{} & \qw & \qw & \qw & \qw & \control{} & \qw & \qw & \qw & \qw & \ctrl{1} & \ctrl{2} & \ctrl{3} & \ctrl{4} & \qw & \qw & \qw & \qw & \qw & \qw & \qw \\
\lstick{\text{Node 4: }$q_{13}$} & \qw & \qw & \qw & \qw & \qw & \qw & \qw & \qw & \qw & \qw & \qw & \control{} & \qw & \qw & \qw & \qw & \qw & \qw & \qw & \qw & \qw & \qw & \qw & \qw & \qw & \control{} & \qw & \qw & \qw & \qw & \qw & \qw & \qw & \qw & \qw & \qw & \qw & \qw & \control{} & \qw & \qw & \qw & \qw & \qw & \qw & \qw & \qw & \qw & \qw & \qw & \control{} & \qw & \qw & \qw & \qw & \qw & \qw & \qw & \qw & \qw & \qw & \control{} & \qw & \qw & \qw & \qw & \qw & \qw & \qw & \qw & \qw & \control{} & \qw & \qw & \qw & \qw & \qw & \qw & \qw & \qw & \control{} & \qw & \qw & \qw & \qw & \qw & \qw & \qw & \control{} & \qw & \qw & \qw & \qw & \qw & \qw & \control{} & \qw & \qw & \qw & \qw & \qw & \control{} & \qw & \qw & \qw & \qw & \control{} & \qw & \qw & \qw & \control{} & \qw & \qw & \qw & \ctrl{1} & \ctrl{2} & \ctrl{3} & \qw & \qw & \qw & \qw \\
\lstick{$q_{14}$} & \qw & \qw & \qw & \qw & \qw & \qw & \qw & \qw & \qw & \qw & \qw & \qw & \control{} & \qw & \qw & \qw & \qw & \qw & \qw & \qw & \qw & \qw & \qw & \qw & \qw & \qw & \control{} & \qw & \qw & \qw & \qw & \qw & \qw & \qw & \qw & \qw & \qw & \qw & \qw & \control{} & \qw & \qw & \qw & \qw & \qw & \qw & \qw & \qw & \qw & \qw & \qw & \control{} & \qw & \qw & \qw & \qw & \qw & \qw & \qw & \qw & \qw & \qw & \control{} & \qw & \qw & \qw & \qw & \qw & \qw & \qw & \qw & \qw & \control{} & \qw & \qw & \qw & \qw & \qw & \qw & \qw & \qw & \control{} & \qw & \qw & \qw & \qw & \qw & \qw & \qw & \control{} & \qw & \qw & \qw & \qw & \qw & \qw & \control{} & \qw & \qw & \qw & \qw & \qw & \control{} & \qw & \qw & \qw & \qw & \control{} & \qw & \qw & \qw & \control{} & \qw & \qw & \control{} & \qw & \qw & \ctrl{1} & \ctrl{2} & \qw & \qw \\
\lstick{$q_{15}$} & \qw & \qw & \qw & \qw & \qw & \qw & \qw & \qw & \qw & \qw & \qw & \qw & \qw & \control{} & \qw & \qw & \qw & \qw & \qw & \qw & \qw & \qw & \qw & \qw & \qw & \qw & \qw & \control{} & \qw & \qw & \qw & \qw & \qw & \qw & \qw & \qw & \qw & \qw & \qw & \qw & \control{} & \qw & \qw & \qw & \qw & \qw & \qw & \qw & \qw & \qw & \qw & \qw & \control{} & \qw & \qw & \qw & \qw & \qw & \qw & \qw & \qw & \qw & \qw & \control{} & \qw & \qw & \qw & \qw & \qw & \qw & \qw & \qw & \qw & \control{} & \qw & \qw & \qw & \qw & \qw & \qw & \qw & \qw & \control{} & \qw & \qw & \qw & \qw & \qw & \qw & \qw & \control{} & \qw & \qw & \qw & \qw & \qw & \qw & \control{} & \qw & \qw & \qw & \qw & \qw & \control{} & \qw & \qw & \qw & \qw & \control{} & \qw & \qw & \qw & \control{} & \qw & \qw & \control{} & \qw & \control{} & \qw & \ctrl{1} & \qw \\
\lstick{$q_{16}$} & \qw & \qw & \qw & \qw & \qw & \qw & \qw & \qw & \qw & \qw & \qw & \qw & \qw & \qw & \control{} & \qw & \qw & \qw & \qw & \qw & \qw & \qw & \qw & \qw & \qw & \qw & \qw & \qw & \control{} & \qw & \qw & \qw & \qw & \qw & \qw & \qw & \qw & \qw & \qw & \qw & \qw & \control{} & \qw & \qw & \qw & \qw & \qw & \qw & \qw & \qw & \qw & \qw & \qw & \control{} & \qw & \qw & \qw & \qw & \qw & \qw & \qw & \qw & \qw & \qw & \control{} & \qw & \qw & \qw & \qw & \qw & \qw & \qw & \qw & \qw & \control{} & \qw & \qw & \qw & \qw & \qw & \qw & \qw & \qw & \control{} & \qw & \qw & \qw & \qw & \qw & \qw & \qw & \control{} & \qw & \qw & \qw & \qw & \qw & \qw & \control{} & \qw & \qw & \qw & \qw & \qw & \control{} & \qw & \qw & \qw & \qw & \control{} & \qw & \qw & \qw & \control{} & \qw & \qw & \control{} & \qw & \control{} & \control{} & \qw \\
    \end{quantikz}%
    }
    \caption{Global gate operation ($GCZ_{16}$) over 4 nodes, with 120 CZs; $\mathrm{GCZ}_{16} \;=\; \prod_{1 \le m < n \le 16} CZ_{q_m,q_n}$}
    \label{fig:gcz120}
\end{figure*}
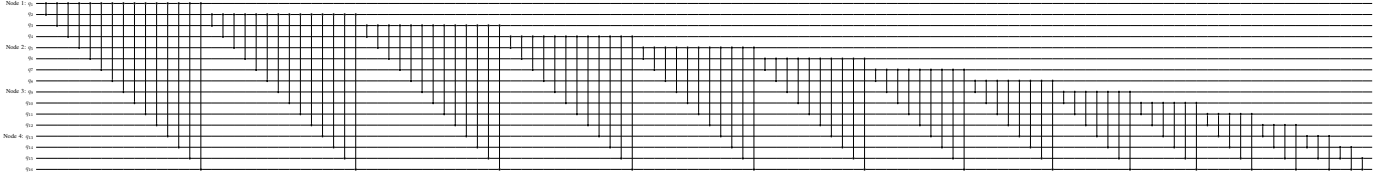

\begin{figure}[t]
    \centering
    \resizebox{\linewidth}{!}{%
\begin{quantikz}[row sep=0.22cm, column sep=0.10cm]
\lstick{\text{Node 1: }$q_1$} & \ctrl{1} & \ctrl{2} & \ctrl{3} & \ctrl{4} & \ctrl{5} & \ctrl{6} & \ctrl{7} & \ctrl{8} & \ctrl{9} & \ctrl{10} & \ctrl{11} & \ctrl{12} & \ctrl{13} & \ctrl{14} & \ctrl{15} & \qw & \qw & \qw & \qw & \qw & \qw & \qw & \qw & \qw & \qw & \qw & \qw & \qw & \qw & \qw & \qw & \qw & \qw & \qw & \qw & \qw & \qw & \qw & \qw & \qw & \qw & \qw & \qw & \qw & \qw & \qw & \qw & \qw & \qw & \qw & \qw & \qw & \qw & \qw & \qw \\
\lstick{$q_{2}$} & \control{} & \qw & \qw & \qw & \qw & \qw & \qw & \qw & \qw & \qw & \qw & \qw & \qw & \qw & \qw & \ctrl{1} & \ctrl{2} & \ctrl{3} & \ctrl{4} & \ctrl{5} & \ctrl{6} & \ctrl{7} & \ctrl{8} & \ctrl{9} & \ctrl{10} & \ctrl{11} & \ctrl{12} & \ctrl{13} & \ctrl{14} & \qw & \qw & \qw & \qw & \qw & \qw & \qw & \qw & \qw & \qw & \qw & \qw & \qw & \qw & \qw & \qw & \qw & \qw & \qw & \qw & \qw & \qw & \qw & \qw & \qw & \qw \\
\lstick{$q_{3}$} & \qw & \control{} & \qw & \qw & \qw & \qw & \qw & \qw & \qw & \qw & \qw & \qw & \qw & \qw & \qw & \control{} & \qw & \qw & \qw & \qw & \qw & \qw & \qw & \qw & \qw & \qw & \qw & \qw & \qw & \ctrl{1} & \ctrl{2} & \ctrl{3} & \ctrl{4} & \ctrl{5} & \ctrl{6} & \ctrl{7} & \ctrl{8} & \ctrl{9} & \ctrl{10} & \ctrl{11} & \ctrl{12} & \ctrl{13} & \qw & \qw & \qw & \qw & \qw & \qw & \qw & \qw & \qw & \qw & \qw & \qw & \qw \\
\lstick{$q_{4}$} & \qw & \qw & \control{} & \qw & \qw & \qw & \qw & \qw & \qw & \qw & \qw & \qw & \qw & \qw & \qw & \qw & \control{} & \qw & \qw & \qw & \qw & \qw & \qw & \qw & \qw & \qw & \qw & \qw & \qw & \control{} & \qw & \qw & \qw & \qw & \qw & \qw & \qw & \qw & \qw & \qw & \qw & \qw & \ctrl{1} & \ctrl{2} & \ctrl{3} & \ctrl{4} & \ctrl{5} & \ctrl{6} & \ctrl{7} & \ctrl{8} & \ctrl{9} & \ctrl{10} & \ctrl{11} & \ctrl{12} & \qw \\
\lstick{\text{Node 2: }$q_5$} & \qw & \qw & \qw & \control{} & \qw & \qw & \qw & \qw & \qw & \qw & \qw & \qw & \qw & \qw & \qw & \qw & \qw & \control{} & \qw & \qw & \qw & \qw & \qw & \qw & \qw & \qw & \qw & \qw & \qw & \qw & \control{} & \qw & \qw & \qw & \qw & \qw & \qw & \qw & \qw & \qw & \qw & \qw & \control{} & \qw & \qw & \qw & \qw & \qw & \qw & \qw & \qw & \qw & \qw & \qw & \qw \\
\lstick{$q_{6}$} & \qw & \qw & \qw & \qw & \control{} & \qw & \qw & \qw & \qw & \qw & \qw & \qw & \qw & \qw & \qw & \qw & \qw & \qw & \control{} & \qw & \qw & \qw & \qw & \qw & \qw & \qw & \qw & \qw & \qw & \qw & \qw & \control{} & \qw & \qw & \qw & \qw & \qw & \qw & \qw & \qw & \qw & \qw & \qw & \control{} & \qw & \qw & \qw & \qw & \qw & \qw & \qw & \qw & \qw & \qw & \qw \\
\lstick{$q_{7}$} & \qw & \qw & \qw & \qw & \qw & \control{} & \qw & \qw & \qw & \qw & \qw & \qw & \qw & \qw & \qw & \qw & \qw & \qw & \qw & \control{} & \qw & \qw & \qw & \qw & \qw & \qw & \qw & \qw & \qw & \qw & \qw & \qw & \control{} & \qw & \qw & \qw & \qw & \qw & \qw & \qw & \qw & \qw & \qw & \qw & \control{} & \qw & \qw & \qw & \qw & \qw & \qw & \qw & \qw & \qw & \qw \\
\lstick{$q_{8}$} & \qw & \qw & \qw & \qw & \qw & \qw & \control{} & \qw & \qw & \qw & \qw & \qw & \qw & \qw & \qw & \qw & \qw & \qw & \qw & \qw & \control{} & \qw & \qw & \qw & \qw & \qw & \qw & \qw & \qw & \qw & \qw & \qw & \qw & \control{} & \qw & \qw & \qw & \qw & \qw & \qw & \qw & \qw & \qw & \qw & \qw & \control{} & \qw & \qw & \qw & \qw & \qw & \qw & \qw & \qw & \qw \\
\lstick{\text{Node 3: }$q_9$} & \qw & \qw & \qw & \qw & \qw & \qw & \qw & \control{} & \qw & \qw & \qw & \qw & \qw & \qw & \qw & \qw & \qw & \qw & \qw & \qw & \qw & \control{} & \qw & \qw & \qw & \qw & \qw & \qw & \qw & \qw & \qw & \qw & \qw & \qw & \control{} & \qw & \qw & \qw & \qw & \qw & \qw & \qw & \qw & \qw & \qw & \qw & \control{} & \qw & \qw & \qw & \qw & \qw & \qw & \qw & \qw \\
\lstick{$q_{10}$} & \qw & \qw & \qw & \qw & \qw & \qw & \qw & \qw & \control{} & \qw & \qw & \qw & \qw & \qw & \qw & \qw & \qw & \qw & \qw & \qw & \qw & \qw & \control{} & \qw & \qw & \qw & \qw & \qw & \qw & \qw & \qw & \qw & \qw & \qw & \qw & \control{} & \qw & \qw & \qw & \qw & \qw & \qw & \qw & \qw & \qw & \qw & \qw & \control{} & \qw & \qw & \qw & \qw & \qw & \qw & \qw \\
\lstick{$q_{11}$} & \qw & \qw & \qw & \qw & \qw & \qw & \qw & \qw & \qw & \control{} & \qw & \qw & \qw & \qw & \qw & \qw & \qw & \qw & \qw & \qw & \qw & \qw & \qw & \control{} & \qw & \qw & \qw & \qw & \qw & \qw & \qw & \qw & \qw & \qw & \qw & \qw & \control{} & \qw & \qw & \qw & \qw & \qw & \qw & \qw & \qw & \qw & \qw & \qw & \control{} & \qw & \qw & \qw & \qw & \qw & \qw \\
\lstick{$q_{12}$} & \qw & \qw & \qw & \qw & \qw & \qw & \qw & \qw & \qw & \qw & \control{} & \qw & \qw & \qw & \qw & \qw & \qw & \qw & \qw & \qw & \qw & \qw & \qw & \qw & \control{} & \qw & \qw & \qw & \qw & \qw & \qw & \qw & \qw & \qw & \qw & \qw & \qw & \control{} & \qw & \qw & \qw & \qw & \qw & \qw & \qw & \qw & \qw & \qw & \qw & \control{} & \qw & \qw & \qw & \qw & \qw \\
\lstick{\text{Node 4: }$q_{13}$} & \qw & \qw & \qw & \qw & \qw & \qw & \qw & \qw & \qw & \qw & \qw & \control{} & \qw & \qw & \qw & \qw & \qw & \qw & \qw & \qw & \qw & \qw & \qw & \qw & \qw & \control{} & \qw & \qw & \qw & \qw & \qw & \qw & \qw & \qw & \qw & \qw & \qw & \qw & \control{} & \qw & \qw & \qw & \qw & \qw & \qw & \qw & \qw & \qw & \qw & \qw & \control{} & \qw & \qw & \qw & \qw \\
\lstick{$q_{14}$} & \qw & \qw & \qw & \qw & \qw & \qw & \qw & \qw & \qw & \qw & \qw & \qw & \control{} & \qw & \qw & \qw & \qw & \qw & \qw & \qw & \qw & \qw & \qw & \qw & \qw & \qw & \control{} & \qw & \qw & \qw & \qw & \qw & \qw & \qw & \qw & \qw & \qw & \qw & \qw & \control{} & \qw & \qw & \qw & \qw & \qw & \qw & \qw & \qw & \qw & \qw & \qw & \control{} & \qw & \qw & \qw \\
\lstick{$q_{15}$} & \qw & \qw & \qw & \qw & \qw & \qw & \qw & \qw & \qw & \qw & \qw & \qw & \qw & \control{} & \qw & \qw & \qw & \qw & \qw & \qw & \qw & \qw & \qw & \qw & \qw & \qw & \qw & \control{} & \qw & \qw & \qw & \qw & \qw & \qw & \qw & \qw & \qw & \qw & \qw & \qw & \control{} & \qw & \qw & \qw & \qw & \qw & \qw & \qw & \qw & \qw & \qw & \qw & \control{} & \qw & \qw \\
\lstick{$q_{16}$} & \qw & \qw & \qw & \qw & \qw & \qw & \qw & \qw & \qw & \qw & \qw & \qw & \qw & \qw & \control{} & \qw & \qw & \qw & \qw & \qw & \qw & \qw & \qw & \qw & \qw & \qw & \qw & \qw & \control{} & \qw & \qw & \qw & \qw & \qw & \qw & \qw & \qw & \qw & \qw & \qw & \qw & \control{} & \qw & \qw & \qw & \qw & \qw & \qw & \qw & \qw & \qw & \qw & \qw & \control{} & \qw \\
\end{quantikz}
}
    \caption{The first 54 CZs, from Figure~\ref{fig:gcz120}.}
    \label{fig:gcz54}
\end{figure}
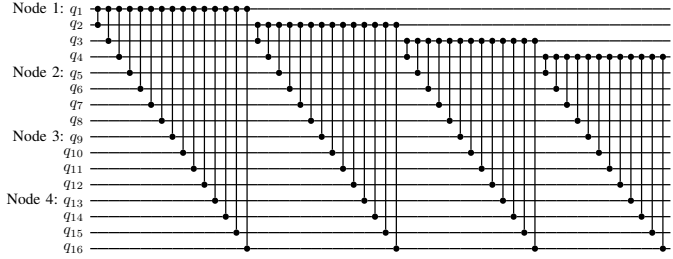

\begin{figure*}[t]
\centering
  \includegraphics[width=1.0\textwidth, height=9cm]{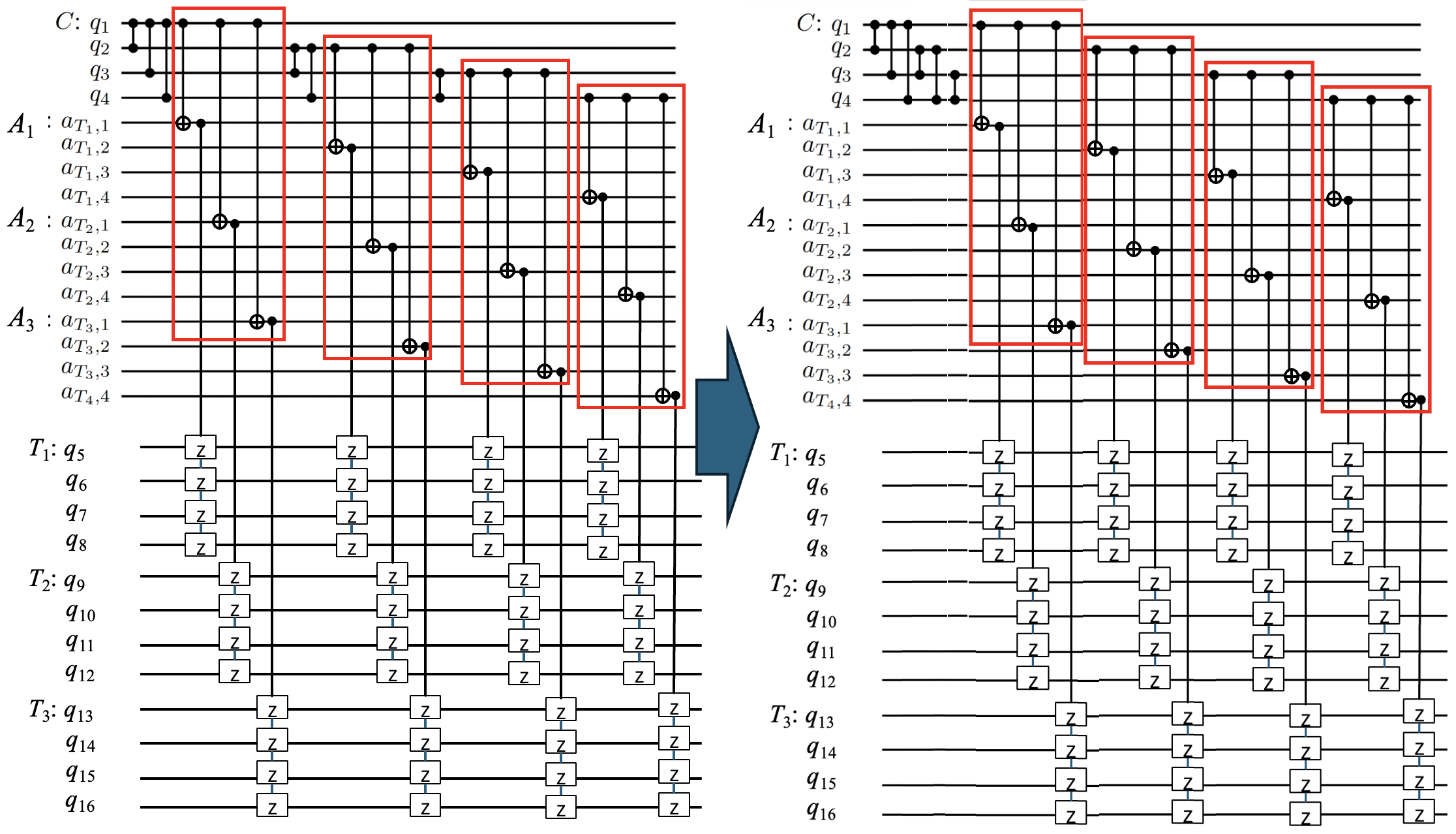} 
  \caption{Partial distributed implementation of the 54 CZs from  Figure~\ref{fig:gcz54} using ancilla qubits, showing (i) reordering with the local CZs ``batching'' and preceding  the rest of the inter-node operations; (ii) the four inter-node fanouts (marked in the four  red boxes) can be implemented via the (distributed version of) transversal $\overline{FANOUT}(C;A_1,A_2,A_3)$ = $\bigotimes_{l \in \{1,...,4\}} \overline{FANOUT}_L(q_l;a_{T_1,l},a_{T_2,l},a_{T_3,l})$. Uncomputation of the ancilla blocks is  needed and shown in Figure~\ref{fig:c-dgcz54}.}
  \label{fig:dgcz54}
\end{figure*}

Figure~\ref{fig:gcz54} shows the first 54 CZs between the (logical) qubits over the four nodes.
Figure~\ref{fig:dgcz54} shows how the first 54 CZs between the logical qubits shown in  Figure~\ref{fig:gcz54} can be implemented in a distributed fashion using ancilla blocks on each of the nodes containing the target blocks. We note that
\begin{itemize}
\item the local CZs (on the node containing the control block) can commute with the preceding non-local (inter-node) CZs (left hand side), and so, can be moved and batched prior to all the inter-node CZs (right hand side);
\item the control logical qubits from the control block $q_1,q_2,q_3,q_4$ are fanned-out to the ancilla qubits on the target nodes - $A_1,A_2$ and $A_3$ are the ancilla logical qubits (or qubit blocks) on target nodes $T_1,T_2$ and $T_3$, respectively - all initialised to $\ket{\overline{0}}_L$;
\item each ancilla qubit is fanned-out to the internal target qubits, within each node.
\end{itemize}

 \begin{figure*}[t]
\centering
  \includegraphics[width=1.0\textwidth]{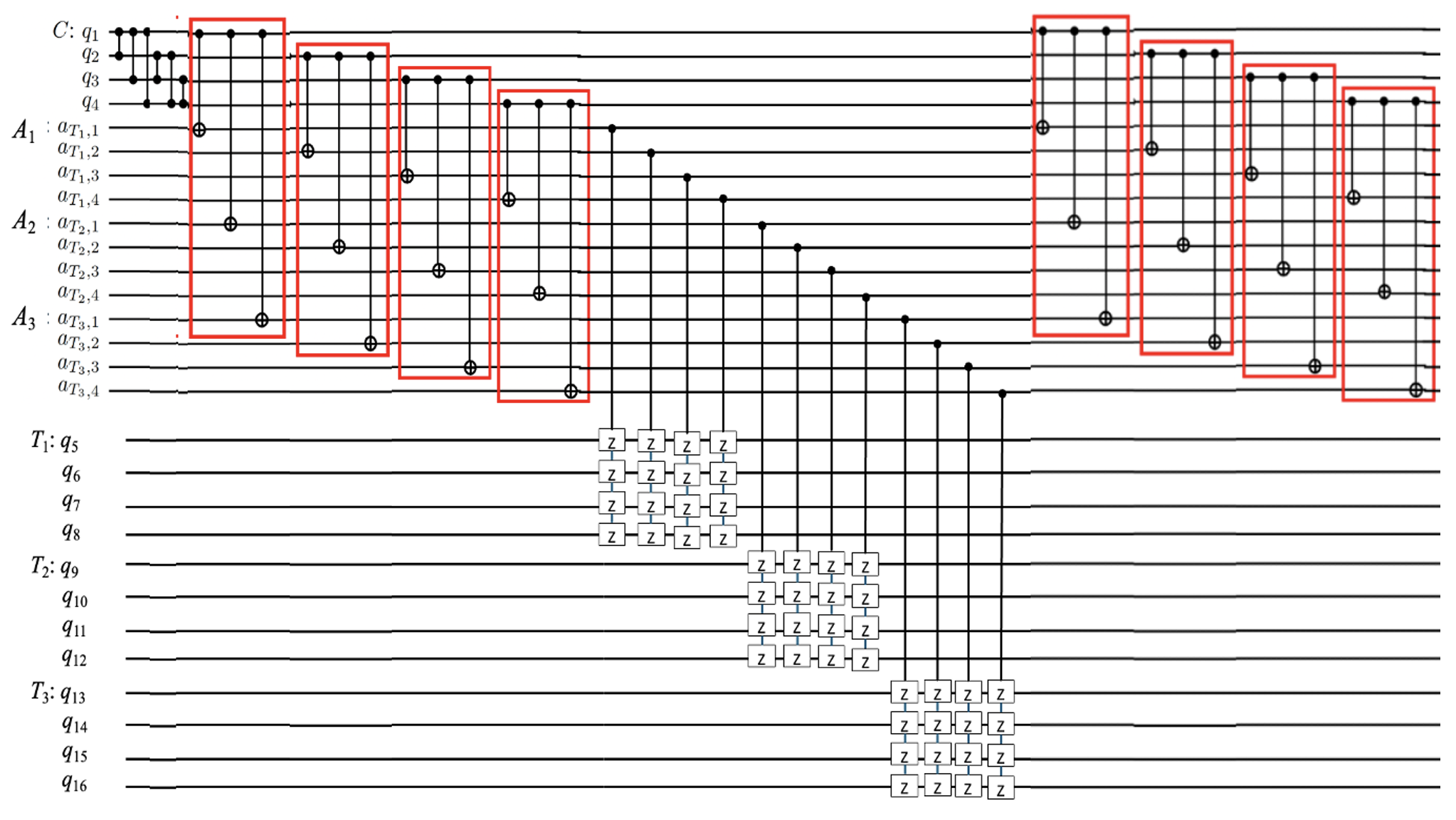} 
  \caption{Starting with clean ancillas set to $\ket{0}_L$, the complete distributed implementation of the 54 logical CZs from  Figure~\ref{fig:gcz54}; the four inter-node fanouts (marked in the four  red boxes) can be implemented via   $d\overline{FANOUT}(C;A_1,A_2,A_3)$ = $\bigotimes_{l \in \{1,...,4\}} d\overline{FANOUT}_L(q_l;a_{T_1,l},a_{T_2,l},a_{T_3,l})$.}
  \label{fig:c-dgcz54}
\end{figure*}

With reordering and uncomputation added to Figure~\ref{fig:dgcz54}, we have the circuit in Figure~\ref{fig:c-dgcz54}.  
Let $C = \{q_1,q_2,q_3,q_4\}$, and 
$T_i =\{q_{4i+1},q_{4i+2},q_{4i+3},q_{4i+4}\}$, for $i=1,2,3$, with ancilla register $A_i = \{a_{T_i,1},a_{T_i,2},a_{T_i,3},a_{T_i,4}\}$ on the same node as $T_i$.
Stage 1: intra‑$C$ CZs; this is effectively a local GCZ operation on node $C$: $GCZ_{C} \;=\; \prod_{1 \le j < k \le 4} CZ_{q_j,q_k}$.
Stage 2: remote CZ block for each node $T_i$: note the concurrent distributed fanouts among the logical qubits on different nodes, the four red boxes in Figure~\ref{fig:c-dgcz54}, which can be realized using the   transversal fanout between the control block and the ancilla blocks, i.e. $\overline{FANOUT}(C;A_1,A_2,A_3)$ = $\bigotimes_{l \in \{1,...,4\}} \overline{FANOUT}_L(q_l;a_{T_1,l},a_{T_2,l},a_{T_3,l})$. We denote by $d\overline{FANOUT}$ the  distributed transversal fanout among blocks on different nodes, i.e. $d\overline{FANOUT}(C;A_1,A_2,A_3)$ = $\bigotimes_{l \in \{1,...,4\}} d\overline{FANOUT}_L(q_l;a_{T_1,l},a_{T_2,l},a_{T_3,l})$

Hence, in terms of fanouts, a distributed version of the 54 CZs, denoted by $dU_{54}$, can be written as a product of local and distributed operations (rightmost operations execute first):
\begin{align*}
 & dU_{54} =  d\overline{FANOUT}(C;A_{1},A_{2},A_{3}) \\
 & \cdot \Pi_{i=1}^{3} \Pi_{j=1}^{4}  \overline{FANOUT}^{Z}_L(a_{T_i,j};q_{4i+1},q_{4i+2},q_{4i+3},q_{4i+4})  \\
 & \cdot d\overline{FANOUT}(C;A_{1},A_{2},A_{3}) \cdot GCZ_{C} 
\end{align*}
Also, note the set of local fanout operations (not CNOTs, but CZs), denoted by $\overline{FANOUT}^{Z}_L$, on each node commute (and so,  can be done concurrently even if depicted in sequence in Figure~\ref{fig:c-dgcz54}). Note that $d\overline{FANOUT}^\dagger=d\overline{FANOUT}$.

Coming back to Figure~\ref{fig:gcz120}, each block of four ``triangles'' can be realized in a similar way, the second block of four using a distributed transversal fanout between a control block in node $T_1$ and nodes $T_2$ and $T_3$, and the last four triangles involve only the nodes $T_1$ and $T_2$, and so needs only (distributed) transversal CNOTs.
In general, with a BB encoding $[[n,k,d]]$, for a $N$ (logical) qubit $GCZ_{N}$ with $k$ logical qubits per node (one block per node), over $m$ nodes, i.e. $N=km$, we require $2(m-1)$ distributed  transversal   $\overline{FANOUT}$ operations (including transversal CNOTs between pairs of nodes), a factor of two due to uncomputation, and  local GCZ and local fanout operations. Also, $n$ physical $m'$-qubit GHZ states (for $m' \leq m$)  are required for each of the  distributed transversal fanouts among distributed code blocks involving $m'$ blocks  (i.e., for the concurrent $k$ logical fanouts among logical qubits over $m'$ nodes). Note that a distributed fanout with one control block and one target block is effectively a non-local CNOT among blocks and instead of GHZ states, we use Bell pairs.  Hence, for a $GCZ_{N=km}$, one $[[n,k,d]] BB$  block per node, $O(nm)$ physical GHZ states (over a varying number of nodes) are required, or for $\frac{n}{k}=c$ for some constant $c$, $O(ckm)=O(km)=O(N)$ physical GHZ states are required.

\section{Conclusion and Future Work}
We demonstrated a distributed transversal fanout for BB-encoded logical qubits. We note that its advantages rely on the ability to form distributed GHZ states efficiently and with low enough noise - our  simulation   demonstrated the advantage of one shot fanout, but other noise models for the GHZ states can be investigated. Also, we showed  how the concurrent logical fanouts from transversal operations can be used to construct distributed implementations of high-fanout CZ/CNOT circuits, including distributed GMS/GCZ constructions.
Future work  could further analyse much larger circuits.

\section*{Acknowledgements}
The author acknowledges the use of ChatGPT-5.6 Luna to assist in (i) generating Python code for the simulation study used in Section~\ref{sec:simulation}, (ii)  generating the diagram in Figure~\ref{fig:gcz120} (the prompt is to generalize from an example of  a manually drawn GCZ with a smaller number of CZ gates), and (iii) generating the Python code for plotting the graphs in Figures~\ref{fig:4BB-LERvsPER}, \ref{fig:10p-4BB-LERvsPER}, \ref{fig:10pandp-8BB-LERvsPER}, and \ref{fig:ler-improvement-8block}. The author verified all algorithmic logic and executed the code, and take full responsibility for the  results.
\bibliographystyle{plain}
\bibliography{refs-new}
\end{document}